\documentclass[conference]{IEEEtran}
\IEEEoverridecommandlockouts
\usepackage{cite}

\usepackage{braket}
\usepackage{amsmath,amssymb,amsfonts}
\usepackage{algorithmic}
\usepackage{graphicx}
\usepackage{textcomp}
\usepackage{xcolor}
\usepackage[ruled,linesnumbered]{algorithm2e} 
\usepackage[nolist]{acronym}
\usepackage{soul}
\usepackage{marginnote}
\usepackage{url}
\usepackage[inline]{enumitem}
\usepackage{tabularx}
\usepackage{amsthm}
\usepackage{tikz}
\usepackage{tikz}
\usepackage{hyperref}
\hypersetup{
    colorlinks=true,
    linkcolor=blue,
    filecolor=magenta,      
    urlcolor=cyan,
    pdftitle={Overleaf Example},
    pdfpagemode=FullScreen,
    }
\usepackage{cleveref}

\newcommand{\hopper}{\textsc{hopper}}
\newcommand{\sync}{\textsc{sync}}

\newcommand{\optional}[1]{%
  {\color{gray}#1%
  }%
}
\renewcommand{\optional}[1]{}

\begin{document}

\title{
  \hopper: A Hop-by-hop Entanglement Distribution Protocol for Asynchronous Quantum Networks
}

\author{
\IEEEauthorblockN{Claudio Cicconetti}
\IEEEauthorblockA{Consiglio Nazionale delle Ricerche -- Istituto di Informatica e Telematica ---
c.cicconetti@iit.cnr.it}
}


\IEEEtitleabstractindextext{%
\begin{abstract}
The quantum Internet relies on the ability to distribute entangled quantum bits (ebits) between quantum memories at the end nodes, to perform applications like blind or distributed quantum computing that are impossible if end nodes are connected via a classical, i.e., non-quantum network. This need creates new challenges due to the fragile nature of entanglement, which decoheres over short timescales and cannot be amplified, buffered, or retransmitted. Two broad categories of approaches have been proposed in the scientific literature to realize such an entanglement distribution in a given path: one relying on a synchronous time-slotted model, and another one where intermediate nodes interact asynchronously. However, both of them implicitly assume a serial operation, where one ebit is established and made available to the application on end nodes before creating a new one. This is inefficient in long-range networks, with high transmission latencies, if the intermediate nodes have multiple memory qubits that could be used in parallel. To overcome this limitation, in this paper, we study the implications of multiplexing concurrent ebit requests on the same quantum, for both synchronous and asynchronous operation. Furthermore, for the latter, we define a novel distribution protocol, called \hopper, where the intermediate nodes make autonomous and hop-by-hop decisions on the use of their local resources when establishing an ebit. With numerical simulations, we show that \hopper is effective in handling multiple ebit requests in parallel, and it exhibits significantly better performance than a synchronous alternative in different scenarios.
\end{abstract}

}

\maketitle

\begin{tikzpicture}[remember picture,overlay]
\node[anchor=south,yshift=10pt] at (current page.south) {\fbox{\parbox{\dimexpr\textwidth-\fboxsep-\fboxrule\relax}{
  \footnotesize{
     \copyright 2026 IEEE.  Personal use of this material is permitted.  Permission from IEEE must be obtained for all other uses, in any current or future media, including reprinting/republishing this material for advertising or promotional purposes, creating new collective works, for resale or redistribution to servers or lists, or reuse of any copyrighted component of this work in other works.
  }
}}};
\end{tikzpicture}%

\IEEEdisplaynontitleabstractindextext

\IEEEpeerreviewmaketitle


\section{Introduction}\label{sec:introduction}

Quantum communication technologies are emerging, providing the building blocks for the creation of quantum networks and, eventually, the future Quantum Internet~\cite{liu_road_2025}.
It is widely recognized that the most important service offered by a quantum network is the distribution of entangled \ac{EPR} pairs of quantum bits (qubits) between end nodes with quantum computing capabilities~\cite{rfc9340}.
These entangled qubits, often called \textit{ebits} in the scientific literature for brevity, unlock new applications that are unfeasible with \textit{classical} (i.e., non-quantum) technologies alone.
For instance: \ac{QKD}, which allows two nodes to generate secret key material without a \ac{PKI} with unconditional security guarantees\footnote{Some QKD protocols, like BB84, rely on a Prepare \& Measure scheme, which does not need entanglement distribution. Such a technology is simpler to realize, from an engineering perspective, than a full-fledged quantum network and is, in fact, already available for production deployment on telco fiber optic infrastructures. However, without entanglement distribution, the property of being ``unconditionally secure'' remains limited to a single point-to-point link, while larger or more complex topologies need trusted nodes which, if compromised, would break end-to-end security.}, blind quantum computing~\cite{de_abreu_towards_2025}, which enables a client to delegate the execution of a batch of quantum operations to a server without the latter having any information about what it is running, anonymous transmission of information~\cite{lipinska_anonymous_2018}, and many others~\cite{rfc9583}.

The fundamental operations of quantum networks have been recently demonstrated independently with different technologies~\cite{kucera_demonstration_2024,delle_donne_operating_2025,almanakly_deterministic_2025}, but only in laboratory or controlled conditions.
These findings suggest an ever-increasing maturity of individual components, though they are not yet ready for commercial operation.
At the moment, it is thus of paramount importance to boost research activities on the ecosystem of architectures, protocols, interfaces, and applications that will materialize the grand vision from the current state of the art scientific results.
One important piece of the puzzle that is still missing is the run-time operation of quantum networks at the link and network layers.
Indeed, most scientific papers use system models based on simplistic assumptions about these layers, and focus on specific aspects to be optimized, such as routing (e.g., \cite{oki_survivable_2026}) and scheduling (e.g., \cite{cicconetti2021request}), or on determining fundamental performance bounds (e.g., \cite{pereira_analytical_2023}), while much less attention has been given to a practical end-to-end operation.

Two archetypes of quantum network architectures were proposed in the literature some years ago, and have been adopted ever since in system models.
In this paper, we refer to these modes as \textit{synchronous}~\cite{pant_routing_2019} vs. \textit{asynchronous}~\cite{kozlowski_designing_2020}.
As the name implies, the synchronous approach requires all the nodes in a network to be synchronized and assumes that all the operations for entanglement distribution occur in time slots of fixed duration, including the local entanglement at the link level and the required classical signaling across intermediate and end nodes (more details on this in \Cref{sec:background}).
This approach is simpler, from an architecture perspective, because the operation can be regulated by a central authority, whose decisions can remain in place for consecutive time slots over a given horizon.
On the contrary, in the asynchronous approach, the nodes follow local procedures reacting to local events generated internally or triggered by neighboring nodes.
Such a system is more flexible than a synchronous one, because it allows different islands in the network to proceed at a different pace, depending on their physical characteristics, and is intended to scale better with large network sizes, since there is no need to propagate classical signaling globally in each time slot.

The contributions of this paper are the following:

\begin{itemize}[parsep=0em,leftmargin=*]
\item We develop a novel protocol (\hopper) for hop-by-hop entanglement from an initiator node to its intended peer that is designed to minimize the overall time required for the operation, thereby limiting the degradation of the qubits involved.
\item We show with simulations that the protocol proposed, despite not relying on any centralized coordination, performs better than an alternative using a synchronous approach and exhibits consistent performance in a broad range of path sizes and network characteristics.
\end{itemize}

The rest of the paper is structured as follows.
In \Cref{sec:background}, we describe the synchronous and asynchronous approaches in the context of our quantum network system model.
The novel contribution, \hopper, which builds on the asynchronous approach, is illustrated in \Cref{sec:hopper}, and then evaluated in \Cref{sec:eval} via extensive simulations.
The most relevant studies in the literature are briefly surveyed in \Cref{sec:soa}.
\Cref{sec:conclusions} summarizes the findings of the paper and future research.

\section{Background}\label{sec:background}

\begin{figure}[tbp!]
\centering
\includegraphics[width=\columnwidth]{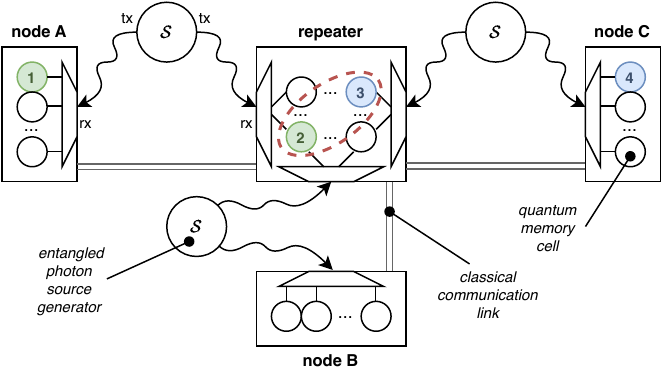}
\vspace{-1.5em}
\caption{Example of a 4-node quantum network to illustrate the system model adopted. There are three end nodes (A, B, and C) and a repeater, all with quantum memories. The quantum links are established via \acp{EPSG} between nodes. To distribute an ebit between A and C, two local EPR pairs are needed: one between A and the repeater (in green) and another between the repeater and C (in blue). To complete the procedure, the repeater then has to perform a \ac{BSM}, followed by \acp{LOCC}.}
\label{fig:network}
\vspace{-1.5em}
\end{figure}

The system model adopted in this contribution is illustrated with the help of the example in \Cref{fig:network}.
The diagram shows a minimal quantum network with three end nodes, labeled from A to C, and a quantum repeater (or \textit{repeater}, for short).
We assume that both end nodes and repeaters have quantum memory cells (\textit{memory cells}, for short) that can hold qubits for some time before they decohere.
The only difference is that repeaters are equipped with multiple quantum \acp{NIC}, which gives them the possibility of extending the distribution of entanglement beyond a single hop.
Likely, end nodes are also equipped with \acp{QPU} to perform quantum network applications, but they are not represented in the diagram because they are outside of the scope of this work.
Two nodes share a quantum communication link if there is an \acf{EPSG} in between that can generate pairs of entangled qubits, which we assume, without loss of generality, to be encoded as photons.
Typically, these entangled qubits are assumed to be Bell states $\ket{\Phi^+} = \frac{1}{\sqrt{2}} \left(\ket{00}+\ket{11}\right)$, which can be used to teleport arbitrary quantum states or implement remote gates for distributed quantum computing\footnote{We do not formally introduce Dirac's notation, i.e., $\ket{~}$, and other quantum information basics, which should not be required to understand the paper's contribution. The interested reader who is not familiar with this topic is referred to, e.g., \cite{rohde_quantum_2025}.}~\cite{barral_review_2025}.
We assume that such a local generation of entangled pairs is \textit{heralded}, i.e., the \ac{EPSG} can detect if the photons emitted from its transmitters (`tx' in the figure) are entangled or not, and can then communicate this information to the nodes involved as classical information~\cite{bernien_heralded_2013}.
Once a photon, i.e., a qubit, reaches a node's receiver (`rx' in the figure), it can be absorbed by a quantum memory cell (or dropped).
In \Cref{fig:network}, there are two entangled pairs $1-2$ and $3-4$, respectively generated between node A and the repeater, and the repeater and node C.
If an ebit between nodes A and C is required, the repeater can obtain this by performing a \acf{BSM}, which involves local operations at the repeater and, crucially, the measurement of the qubits 2 and 3.
Measurement operations in quantum information technologies are destructive, as they collapse the quantum states of qubits to produce classical bits instead.
Such a BSM then empties the memory cells previously occupied by 2 and 3 at the repeater, but forces a remote interaction between the qubits occupying the memory cells 1 and 4, on nodes A and C, respectively.
The end-to-end entanglement process is complete after node A (or node C) has performed some \acf{LOCC}, i.e., the correction of a possible bit/phase flip induced by the measurements at the repeater, depending on the classical bits produced, via X/Z gates.
This process, called \textit{entanglement swapping}, can be reiterated to create ebits between nodes separated by an arbitrary number of repeaters.

\begin{figure}[tbp!]
\centering
\includegraphics[width=\columnwidth]{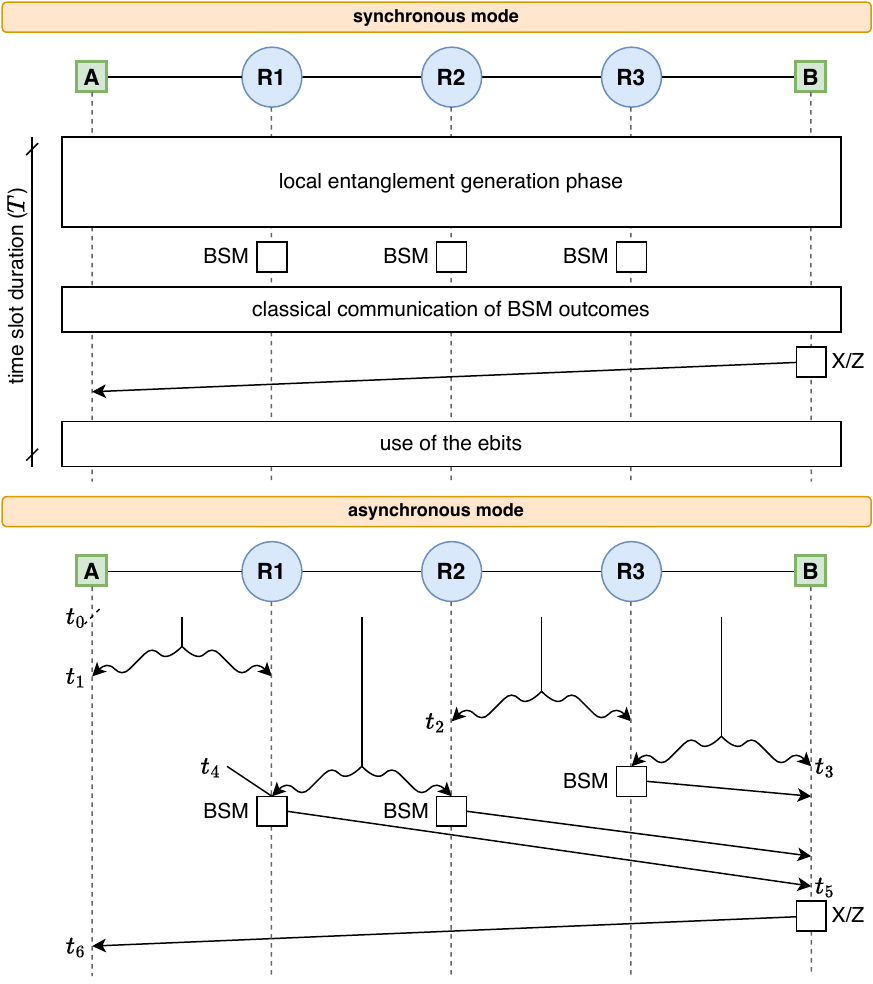}
\vspace{-1.5em}
\caption{Comparison of synchronous (top) vs.\ asynchronous (bottom) approaches, in an example path of three repeaters (R1, R2, and R3) collaborating to distribute ebits between end-nodes A and B. The asynchronous sequence diagram is a modified version of Fig.~6 in \cite{kozlowski_designing_2020}, not showing the path reservation messages and with simplified signaling for the distribution of the \ac{BSM} outcomes.}
\label{fig:mode}
\vspace{-1.5em}
\end{figure}

From what we have described above, it follows that some coordination is required between the nodes in a quantum network to create the ebits, for which, today, there are no practical implementations or standardized protocols.
As briefly introduced in \Cref{sec:introduction}, in the literature, two approaches have been proposed, which we now illustrate, with necessary adaptations to fit our system model and terminology.
A \textit{synchronous} approach can be used, as originally proposed in \cite{pant_routing_2019} and illustrated in the top part of \Cref{fig:mode}, which shows a single path with three repeaters between the two end-nodes.
In this operation mode, the time is divided into slots of fixed duration, which are the same for all the nodes in the network.
A time slot consists of the following phases, each with a specified duration.
First, entangled \ac{EPR} pairs are generated at all the network links (A-R1, R1-R2, etc.).
If heralded entanglement generation, which is a stochastic process, does not complete within the phase duration on all the links, then an ebit cannot be established in this time slot.
Otherwise, all the repeaters perform a \ac{BSM} at the same time, each producing classical bits that are transmitted to an end-node, e.g., B, which performs X/Z corrections and notifies its peer end-node of the failure vs.\ successful creation of the ebit.
In case of success, the end-points can then use the end-to-end entangled qubits at their respective memories until the beginning of the next time slot.
Intuitively, the time slot duration ($T$) should be sufficient to
\begin{enumerate*}[label=(\roman*)]
  \item guarantee a sufficiently high probability that all the local entanglements will be completed on time, and
  \item propagate classical signal from all the repeaters to one of the end points and then to the other one, i.e., which requires 1~\ac{RTT}.
\end{enumerate*}
Since the ebit rate is limited by $1/T$, a long time slot can be a factor dramatically limiting the performance in large and heterogeneous networks, since every entanglement distribution is forced to keep the pace of the slowest and longest possible path.

To cope with such a shortcoming, an asynchronous approach has been proposed as part of the \ac{QNP}~\cite{kozlowski_designing_2020}, illustrated in the sequence diagram in the bottom part of \Cref{fig:mode}.
As can be seen, in this case, there are no time slot boundaries and nodes are not required to make synchronized actions\footnote{A tight time synchronization will likely be needed for the heralded entanglement generation procedure, but this can remain local at every \ac{EPSG}/link.}.
Assuming that at $t_0$ the nodes have preliminarily agreed to establish an ebit between A and B along the path R1--R2--R3, local entanglement is repeatedly attempted at every link.
In the example, A--R1 is the first successful local entanglement, at $t_1$, resulting in the nodes A and R1 keeping their qubits in their respective quantum memories.
Then, each repeater attempts a \ac{BSM} as soon as it receives both the qubits on its left and right link.
In the example, this first happens at $t_3$ for R3, which then sends the measurement outcomes to (e.g.) end-node B.
This is done by R1 and R2 at $t_4$, resulting in the end-node B, at $t_5$, being able to perform the X/Z corrections needed to ensure proper configuration of the ebit, which is communicated to end-node A at $t_6$.
This mode is more complex than the synchronous one but may lead to better performance:
\begin{enumerate*}[label=(\roman*)]
  \item there is no time wasted in the local entanglement generation phase, which advances as soon as possible with no preconfigured duration, and
  \item every path automatically keeps its own pace, i.e., faster/shorter paths will exhibit a higher ebit rate than that of slower/longer ones.
\end{enumerate*}

Many studies in the scientific literature have referred to the synchronous or asynchronous mode as part of their system models while addressing some specific research challenge.
However, to the best of our knowledge, the following questions regarding a possible practical implementation have not received much attention so far:

\begin{enumerate}[parsep=0em,leftmargin=*,label=\arabic*.]
  \item How to multiplex multiple concurrent ebit requests in a network, in particular when there are (partially) overlapping paths?
  \item How to reduce signalling between nodes, to reduce the performance degradation due to decoherence in quantum memories?
  \item What is the quantitative impact on performance of adopting a synchronous approach, compared to an asynchronous equivalent in the same conditions?  
\end{enumerate}

We address the first two questions in the following section, while the last one will be tackled in \Cref{sec:eval}.

\section{\hopper: a hop-by-hop entanglement distribution protocol}\label{sec:hopper}

In this section, we illustrate \hopper, our proposed framework to establish ebits in a quantum network, defined according to the system model in \Cref{sec:background}.
The key consideration we make is that, for a quantum network protocol, \textit{time is of the essence}: due to the necessity of using microscopic physical platforms, showing quantum mechanical effects, to realize qubits, they are much more prone to degradation due to several environmental factors, such as temperature changes and radiation, than macroscopic components used by today's electronic devices.
While there are ongoing commendable efforts towards achieving so-called \ac{QEC}, e.g., with bosonic logical qubits \cite{putterman_hardware-efficient_2025}, there is a wide consensus in the research community that quantum technologies will remain ``noisy'' in the medium-term.
The requirement of reducing classical signaling translates into the nodes making decisions as independently as possible.
For this reason, we select an asynchronous mode of operation for \hopper; such a choice is substantiated with simulations in \Cref{sec:eval}, showing superior performance than with synchronous operation.

\begin{figure}[tbp!]
\centering
\includegraphics[width=\columnwidth]{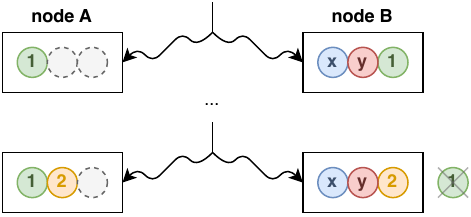}
\vspace{-1.5em}
\caption{Example of inconsistency of quantum memories at two nodes, A and B, due to independent decisions made by them. Different entangled \ac{EPR} pairs have different colors.}
\label{fig:example-memory}
\vspace{-1.5em}
\end{figure}

Making independent decisions at nodes clashes with the nature of ``entanglement'', which is inherently distributed between two qubits.
Let us consider the simple example in \Cref{fig:example-memory}, where two nodes A and B, with 3-qubit quantum memories, share a link to receive entangled \ac{EPR} pairs by an \ac{EPSG}.
When the \ac{EPR} pair number 1 arrives at A and B, it is saved into the first cell at node A, whose other cells are unoccupied, but into the third cell at node B, while the other two cells are busy with \ac{EPR} pairs associated with other links.
If none of the qubits at B are used before the next \ac{EPR} pair 2 arrives, then B will have to overwrite an existing qubit, e.g., number 1 in the example.
This leads to an inconsistency with node A, which instead has room for the new \ac{EPR} pair.
Any other choice, e.g., overwriting another \ac{EPR} pair or just dropping the incoming \ac{EPR} pair 2, would lead to some equivalent inconsistency.
Existing papers do not consider this case, or they assume that some negotiation protocol will be employed to resolve the inconsistency, which, however, would require time, during which the qubits cannot be used while they decohere.

\begin{figure}[tbp!]
\centering
\includegraphics[width=\columnwidth]{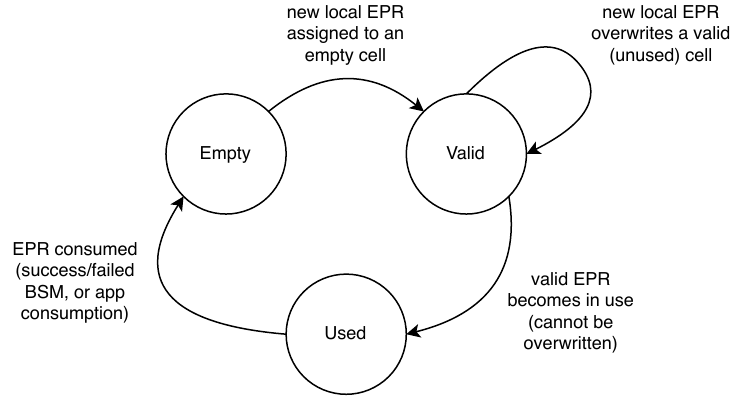}
\vspace{-1.5em}
\caption{\acf{FSM} of a quantum memory cell.}
\label{fig:fsm}
\vspace{-1.5em}
\end{figure}

We solve the inconsistency by allocating quantum memory cells to links, inspired by what was proposed in~\cite{soon_performance_2024}.
However, we assume here that links are directed, with the direction given by the relative position in the path from the \textit{source} end-node, i.e., the one that initiates the procedure of ebit establishment, to the \textit{destination}, i.e., the node that accepts to take part in the end-to-end entanglement.
The discovery of services available at the nodes is outside the scope of this work, like all the signaling needed for the negotiation of capabilities, authorization, and authentication, etc.
At each node, we associate each link with two sub-groups of the quantum memory cells, labeled with the roles master or slave.
For every two nodes $u,v$ sharing a link, every master memory cell at a node $u$ for that link has a mirror slave memory cell on $v$, and the other way around.
The \textit{master} memory cells are under the control of the node itself, which can choose to associate them with ongoing ebit establishment procedures.
On the other hand, a node never takes the initiative in using its \textit{slave} memory cells, which instead are reserved for operations decided by the master.
More specifically, the \ac{FSM} of every memory cell is illustrated in \Cref{fig:fsm}.
When a new qubit arrives at a node on a given link for one role, it can be absorbed in an Empty cell, if available, or it can overwrite a memory cell with a Valid qubit, if it is not being currently used by an ongoing ebit establishment procedure.
In this case, since qubits decohere over time, the straightforward design choice is to always overwrite the oldest Valid memory cell.
If there is no Empty or Valid memory cell, the incoming qubit must be dropped.
This decision is made independently by all the nodes, adopting the same strategy.
On the contrary, a transition from Valid to Used, which effectively \textit{locks} the memory cell to a specific ebit being established, is made only by the node with a master role and accepted by the one with a slave role.
Crucially, when a node overwrites a Valid memory cell, it does not have to communicate this to its peer; therefore, the protocol does not provide any strong consistency guarantees, which are traded for reduced signaling.

\begin{figure*}[tbp!]
\centering
\includegraphics[scale=0.7]{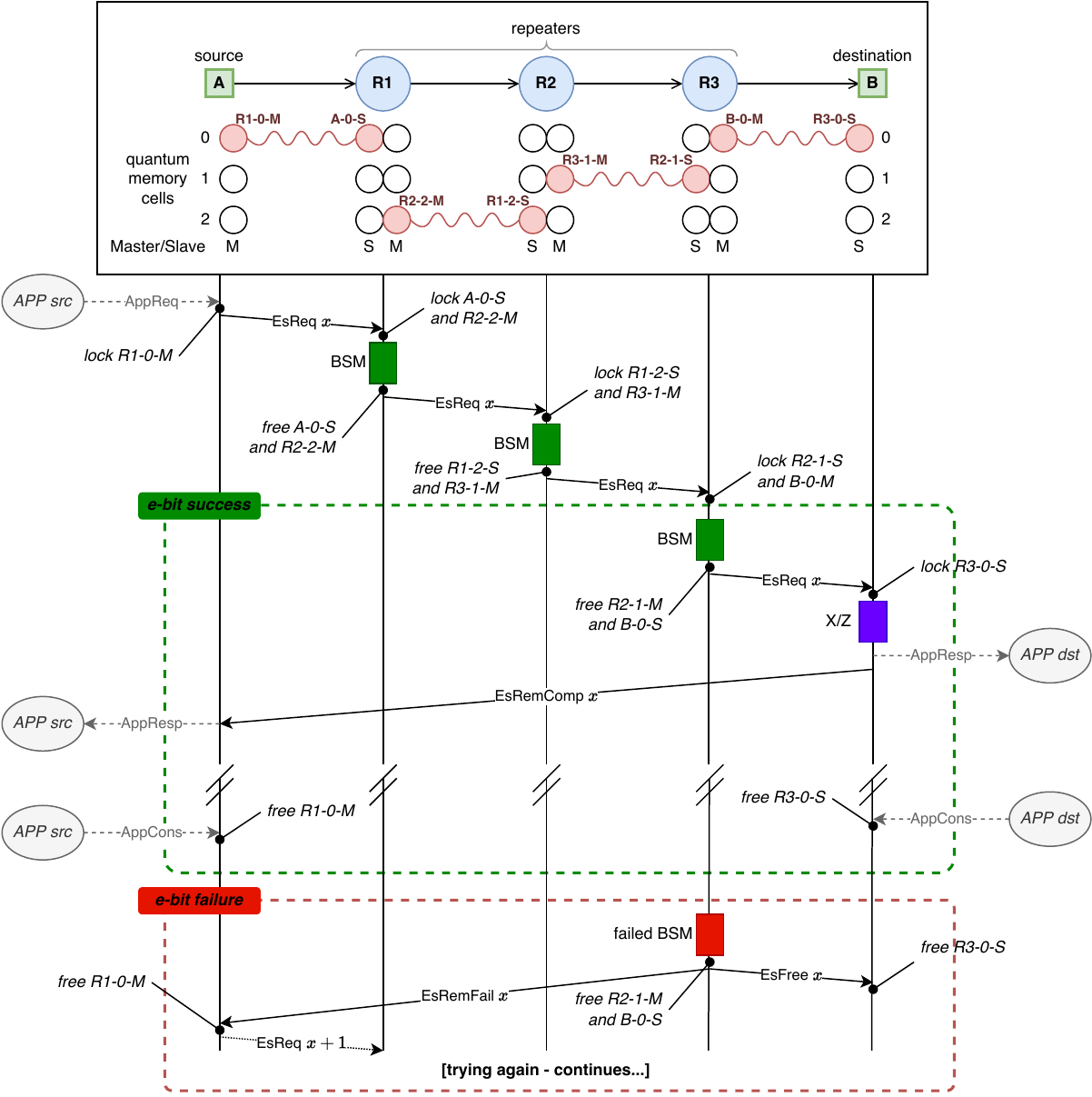}
\vspace{-0.5em}
\caption{Sequence diagram of the procedure to create an ebit between end-nodes A and B via intermediate nodes R1, R2, and R3, all with three memory cells assigned to the links in the path from A to B. A memory cell associated with a link is identified as a triplet $n$-$i$-$r$, where $n$ is the node identifier, $i$ is a 0-based index, and $r$ is the role (Master `M' or Slave `S'); such a notation is only for illustration purposes in the diagram, while in the signaling messages, numerical identifiers are used. The sequence diagram shows two possible courses of action: the ebit is successfully established and used by the applications on the end-nodes (green dashed box) vs.\ failure to create the ebit due to the \ac{BSM} at R3 (red dashed box), resulting in a subsequent retry from A (not shown). In addition to the classical messages exchanged (EsReq, EsRemComp, EsRemFail), the diagram also shows, in gray, internal system calls to inform the applications (AppReq, AppResp), as well as the lock and free operations on the quantum memory cells.}
\label{fig:sequence}
\end{figure*}

Let us now illustrate in detail the proposed ebit establishment procedure of \hopper, with the help of the sequence diagram in \Cref{fig:sequence}, showing an ebit establishment procedure between two nodes, A and B, via three intermediate repeaters, R1, R2, and R3.
For simplicity of illustration, all the links are assumed to have three memory cells.
The procedure begins with an application \textit{src} on the source node A requesting an ebit with a peer application \textit{dst} on the destination node B.
We assume this request, called AppReq, to be passed to the node's operating system, which will take charge of it and notify the application when the ebit is established.
In practice, AppReq might include performance guarantees, such as minimum fidelity, or constraints, such as a timeout after which the operation fails if the ebit is not established, but we do not explore these details here to remain focused on the basic establishment procedure.
For simplicity of illustration, we assume that the path is decided by the source node, based on complete knowledge of the topology, and recorded in the signaling messages.
Alternative design options are feasible, e.g., by considering hierarchical routing inspired by the classical Internet~\cite{jordan-parra_position_2024}, the basic requirement being that there is a way for each node to know what its successor is towards the destination.

End-to-end entanglement is created along the path by means of classical messages of type EsReq, which are sent hop by hop from the source to the destination.
EsReq must contain sufficient information to identify the applications in the network and distinguish between multiple ebits of the same application.
Inspiring by TCP/IP, for this purpose we could use a \textit{five-tuple}  containing the source node identifier and port, the destination node identifier and port, and a unique identifier of the ebit creation attempt, which can be a simple counter incremented at the source node.
The source node is free to select the Valid memory cell it prefers, which is locked, i.e., moves to an Used state in the FSM (see again \Cref{fig:fsm}) to guarantee that new qubits arriving from the \ac{EPSG} will not overwrite it.
By choosing a memory cell, in its Master role, the recipient ---R1 in the example--- will be forced to accept this decision and assign the corresponding Slave memory cell to the ebit under preparation.
When a node receives an EsReq and is not the intended destination, it accepts the decision made by its predecessor on which memory cell is assigned to the ebit, and it makes a decision by itself, selecting a Master memory cell from the pool available at the link towards its successor.
The qubits in the two memory cells, e.g., those identified as A-0-S and R2-2-M in the example for node R1, are then subject to a \ac{BSM} to perform entanglement swapping.
As explained in \Cref{sec:background}, this operation consumes the qubits involved, which can then be freed, i.e., moved to an Empty state, while producing two classical bits, which are embedded in the next EsReq message.
Thus, the end node will receive the full list of results, which allows it to perform the required local quantum operations for X/Z corrections, after which it can inform the \textit{dst} application at the same node, and the source node A via an EsRemComp message, signaling the successful completion of the ebit creation.
There are, however, several reasons why intermediate steps may fail.
For instance, the \ac{BSM} operation may fail, which invalidates the entire chain of entanglement swappings so far.
This is shown in the ``ebit failure'' box in \Cref{fig:sequence}, where node R3 subsequently sends an EsFree message to its successor to free the qubit occupied by the slave memory cell involved at the BSM, and an EsRemFail message to the source to notify that the ebit could not be established and, after freeing the initial memory cell, it should try again.
Or, when an EsReq arrives at a node with an indication of a given memory cell to be used, the latter has already been overwritten by a new qubit that arrived from the \ac{EPSG}; this can be detected by including an identifier for each \ac{EPR} pair generated, which can be a simple counter increased by both nodes when receiving the qubits.
This case is not illustrated in the example due to limited space.
In the simulation tool used for the performance evaluation in \Cref{sec:eval}, we have captured all the possible cases, which provides a validation of the protocol proposed.

In summary, the key design innovations of our framework are the following:
\begin{enumerate*}[label=(\roman*)]
  \item each ebit is assigned an identifier, which is used together with node/port identifiers to uniquely address it in the network;
  \item end-to-end entanglement is realized hop by hop by performing entanglement swapping at intermediate nodes from a source to a destination, embedding the \ac{BSM} classical outputs along the way ---this ensures that an ebit can be created within a single \ac{RTT} without any previous arrangements; and,
  \item nodes use a Master vs.\ Slave role to keep the consistency of \acp{FSM} for their memory cells without explicit negotiation, which reduces latency at the cost of possible failures.
\end{enumerate*}

\subsection{Limitations}

The framework illustrated has some limitations, reported in the following, which we have not addressed so far but are currently under investigation.
First, the problem of an optimal partitioning of the memory cells available at nodes into groups per link/role is not trivial.
In the simulation tool developed and used to obtain the results in \Cref{sec:eval}, we have implemented a simple greedy strategy aimed at balancing the memory cells evenly across the possible paths between end nodes, assuming a shortest path selection.
However, more sophisticated policies could be studied, even dynamic ones, possibly enhancing the performance under specific conditions, especially with unbalanced/heterogeneous demands.
Second, we do not consider purification, which is the \ac{LOCC} process that can increase the quality of a pair of entangled qubits by sacrificing other \ac{EPR} pairs.
This is particularly interesting when considering the case of new qubits generated when there are no Empty cells, in which case one could use the qubits in a Valid memory cell for purification, instead of dropping them as illustrated above.
Performance bounds of such a process have been found in~\cite{inesta_entanglement_2026} under some assumptions.
However, purification requires signaling, which increases latencies, hence decoherence, potentially canceling the benefits, so an in-depth investigation would be needed to quantify the performance trade-offs.
Finally, we have considered only bipartite entanglement, which is the entanglement between two qubits/nodes.
Many quantum applications of practical interest, such as secret voting~\cite{liu_efficient_2011}, require multiple ($> 2$) parties to share entanglement, e.g., via a \ac{GHZ} state $\frac{1}{\sqrt{2}}\left(\ket{00\ldots0} + \ket{11\ldots1} \right)$.
This can be achieved by collating an appropriate number of bi-partite states, as allowed by \hopper, but a more efficient way would be to distribute the multi-partite state directly across the nodes, which is theoretically feasible even if challenging from an engineering perspective~\cite{wu_generation_2022}.
For that, non-trivial modifications would be needed to \hopper, since there would be no ``path'' from source to destination but a full-mesh of peers, instead.

\section{Performance evaluation}\label{sec:eval}

In this section, we evaluate the performance of \hopper, previously illustrated in \Cref{sec:hopper}, compared to a synchronous protocol (see \Cref{sec:background}, called \sync{} below) running in the same conditions.
We focus on two key metrics: the \textit{throughput}, i.e., the number of successful established ebits per second, and the \textit{fidelity}.
For the latter, we do not provide a formal definition, which would require introducing notation and concepts that have been left aside in the paper for brevity.
Intuitively, the fidelity is a scalar number in $[0,1]$ associated with two qubits that measures how close they are to one another, where 1 means that they are identical; a value of 0.5 represents the minimum value for two qubits exhibiting quantum entanglement.
Different quantum network applications require different fidelity thresholds to run properly.
\acp{EPSG} are assumed to generate heralded entangled \ac{EPR} pairs randomly following a Poisson process, with an average rate.
The initial fidelity of local \ac{EPR} pairs is constant, but the end-to-end fidelity is reduced due to two reasons:

\begin{enumerate}[parsep=0em,leftmargin=*,label=\arabic*.]
  \item Dephasing decreases the fidelity over time as $F(t)=\frac{1}{4} + \left(F_{init} - \frac{1}{4}\right) e^{-\Gamma t}$, where $\Gamma$ is the decay rate~\cite{inesta_entanglement_2026}.
  \item The fidelity decreases at every entanglement swapping. If $F_1$ and $F_2$ are the fidelities of the two qubits taking part in the \ac{BSM}, the result is (assuming $p_1=p_2=\eta$ in \cite{greenberger_quantum_1999}): $F = \frac{1}{4} + \frac{3}{4}\left(\frac{4F_1-1}{3}\right)\left(\frac{4F_2-1}{3}\right)$.
\end{enumerate}

\begin{table}
  \centering
  \vspace{-1.5em}
\caption{Main simulation parameters.}\label{tab:sim-params}
  \begin{tabular}{ll|ll}
  \hline
  \textbf{Parameter} & \textbf{Value} & \textbf{Parameter} & \textbf{Value}\\ \hline
  Simulation duration & 60 s & Initial fidelity & 0.95 \\
  BSM probability & 0.95 & \ac{EPSG} rate & 100 Hz \\
  BSM duration & 1 ms & Confidence interval & 95\% \\
  X/Z operations duration & 1 ms & Num replications & 10 \\
  \hline
  \end{tabular}
  \vspace{-1.5em}
\end{table}

The event-driven simulator used was developed in the Rust programming language and is publicly available on GitHub\footnote{\url{https://github.com/ccicconetti/qnet_sim}} with an MIT license, together with the scripts to run the experiments and post-process the results, to ensure full reproducibility.
The physical distance between nodes, decay rate ($\Gamma$), and number of memory cells per node depend on the specific experiment.
Other relevant simulation parameters are reported in \Cref{tab:sim-params}.
The simulator allows arbitrary network topologies, even if in this paper we only study linear chains, a.k.a. individual paths, for a reasonable comparison with the synchronous protocol, and due to limited page budget.

\begin{figure}[tbp!]
\centering
\includegraphics[width=\columnwidth]{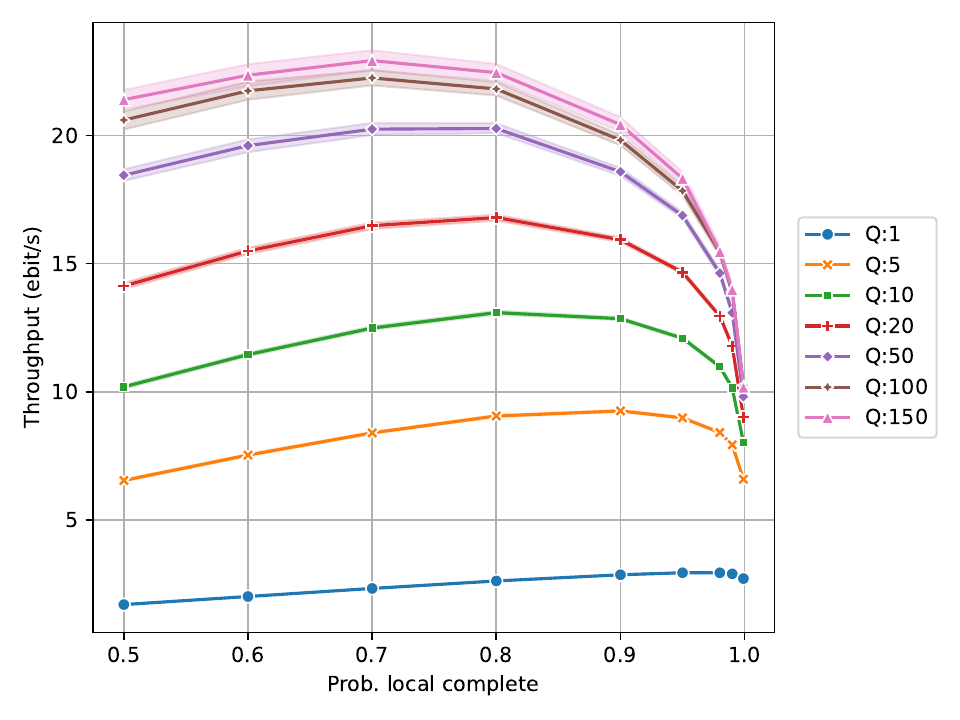}
\vspace{-1.5em}
\caption{Long-distance regime. Throughput with \sync{}, with varying probability of successfully completing the local entanglement phase and quantum memory size (Q).}
\label{fig:005-chain-mini-scalar-throughput}
\vspace{-1.5em}
\end{figure}

We begin our analysis with a linear chain with three repeaters, same as illustrated in the examples in \Cref{sec:background} and \Cref{sec:hopper}, with a varying number of memory cells up to 150, and a \textit{long-distance regime}, where each node receives \ac{EPR} pairs from an \ac{EPSG} 5,000~km away, which is compatible with a satellite \ac{LEO} constellation~\cite{chen_integrated_2021}.
For consistency with the high latencies incurred, we have assumed $\Gamma = 1$~Hz.
We consider \sync{} first, whose performance also depends on a system parameter, i.e., the duration of the local entanglement phase: the longer this duration, the lower the number of ebit attempts, but they will succeed with a higher probability than with a shorter local entanglement phase duration.
This trade-off is explored with simulations in \Cref{fig:005-chain-mini-scalar-throughput}, where the local entanglement phase is expressed indirectly on the $x$-axis through the probability that a local entanglement completes successfully.
As can be seen, the curves show maxima corresponding to different probabilities of local entanglement completion for different quantum memory sizes.
In this study, we do not delve further into the matter, but consider in the following comparison with \hopper{} the probability of local entanglement completion that always provides the highest throughput, giving \sync{} an impractical advantage.
Note that \sync{} is also provided with another unfair advantage: in practice, it would require a preliminary phase to open the path between source and destination, which is not simulated, while \hopper{} can naturally establish ebits between any two nodes through its hop-by-hop protocol.

\begin{figure}[tbp!]
\centering
\includegraphics[width=\columnwidth]{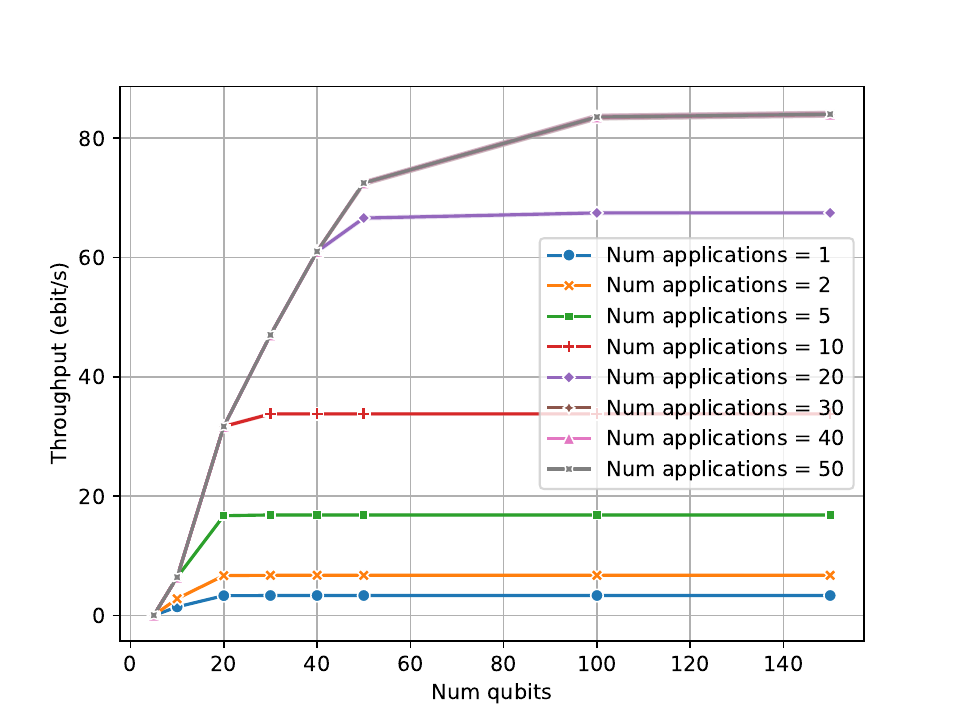}
\vspace{-1.5em}
\caption{Long-distance regime. Throughput with \hopper, with an increasing number of applications, and quantum memory size. The curves with 30, 40, and 50 applications are fully overlapping with one another.}
\label{fig:004-chain-perfect-scalar-throughput}
\vspace{-1.5em}
\end{figure}

As shown in \Cref{fig:004-chain-perfect-scalar-throughput}, the throughput of \hopper{} increases with both the quantum memory size, because of the higher amount of resources, and the number of concurrent applications, because of the higher statistical multiplex that can better keep the ``pipe'' full.
For every number of applications, the plot shows a memory size that allows the throughput to reach a saturation point.
Such a size is smaller with fewer applications.
For instance, with a single application, the throughput saturates already with 20-qubit memories, whereas with 20 applications it does so with 50-qubit memories.
It is interesting to note that with more than 30 applications, there is no further increase in throughput, irrespective of the memory size, because of another choke point due to the \ac{EPSG} rate.

\begin{figure*}[tbp!]
\centering
\includegraphics[width=\columnwidth]{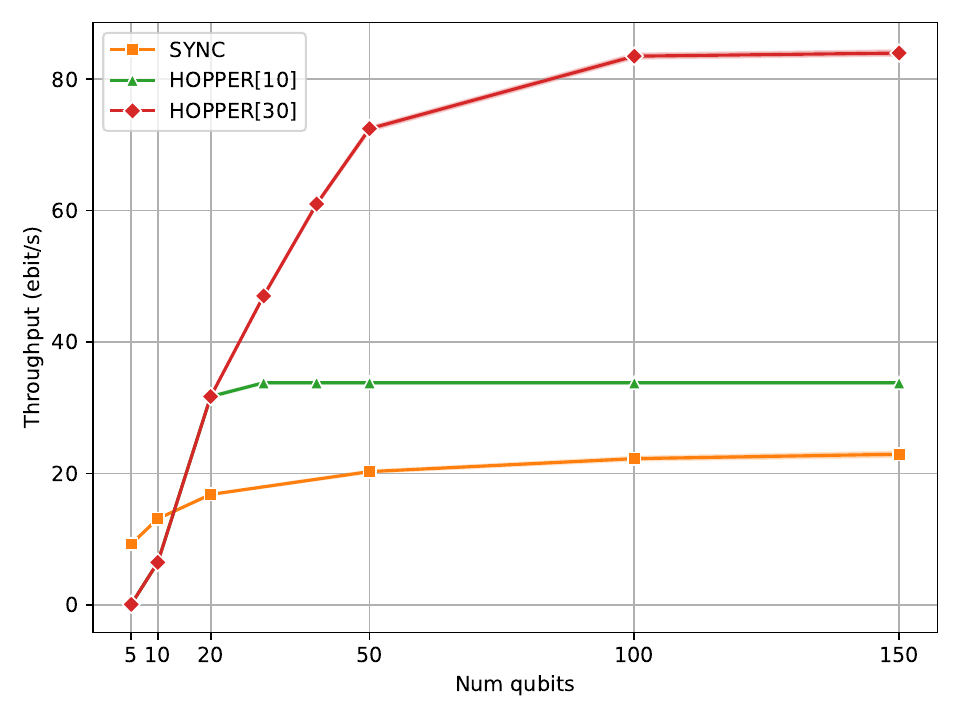}
\includegraphics[width=\columnwidth]{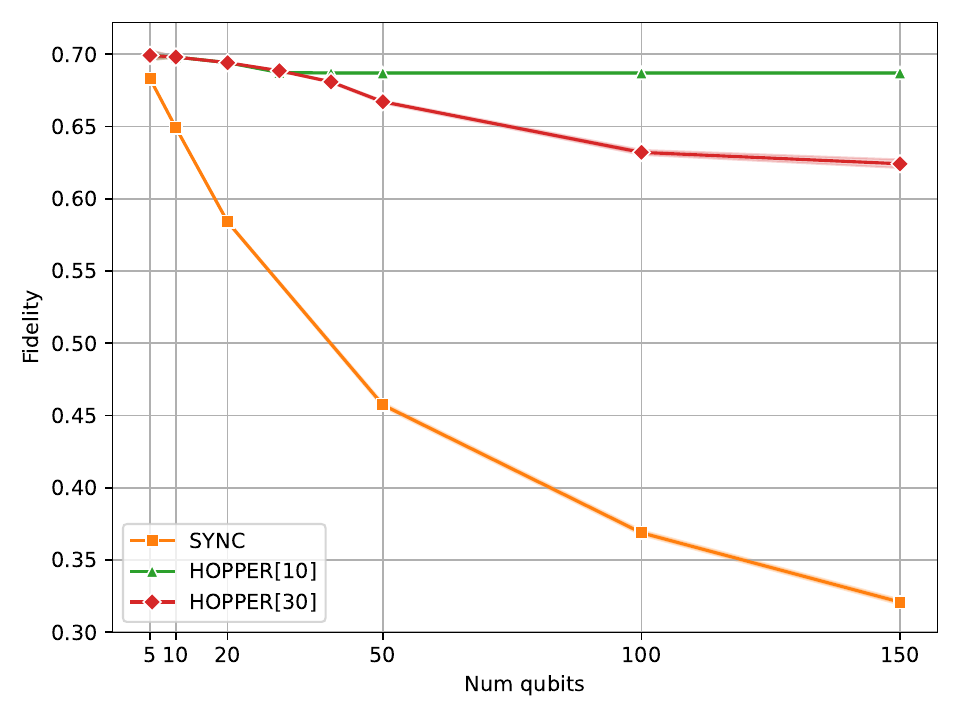}
\vspace{-1.5em}
\caption{Long-distance regime. Comparison of throughput (left) and fidelity (right) between \sync{} and \hopper, the latter when 10 or 30 concurrent ebits procedures are running.}
\label{fig:005-chain-mini-compare}
\end{figure*}

In \Cref{fig:005-chain-mini-compare} we compare the throughput between \sync{} and \hopper~(with 10 and 30 applications).
With a very small number of qubits, \hopper's throughput is vanishing because the protocol consistently runs into the case where a Valid memory cell is overwritten by a node with Slave role before the Master can lock it.
However, this condition disappears with a sufficient number of qubits in the memory, depending on the classical signaling latency, i.e., the distance between the Master and Slave node.
With more than 10 qubits, \hopper's throughput increases fast, then saturates to a value of about 84 ebits/s, with 30 applications, which is remarkably close to the link rate (100~\ac{EPR} pairs/s).
With 10 applications, the throughput reaches a lower plateau at the same number of qubits.
Instead, \sync{} increases more slowly and reaches a throughput of little more than 20 ebits/s.
When looking at the fidelity, we find that it decreases with all the approaches as the memory size increases, because the higher throughput is gained at the cost of multiplexing concurrent ebit procedures by keeping qubits longer in the memories.
However, the \sync{} curve decreases much faster than those for \hopper, which shows that the latter has better throughput \textit{and} fidelity due to a wiser utilization of the resources available.

\begin{figure}[tbp!]
\centering
\includegraphics[width=\columnwidth]{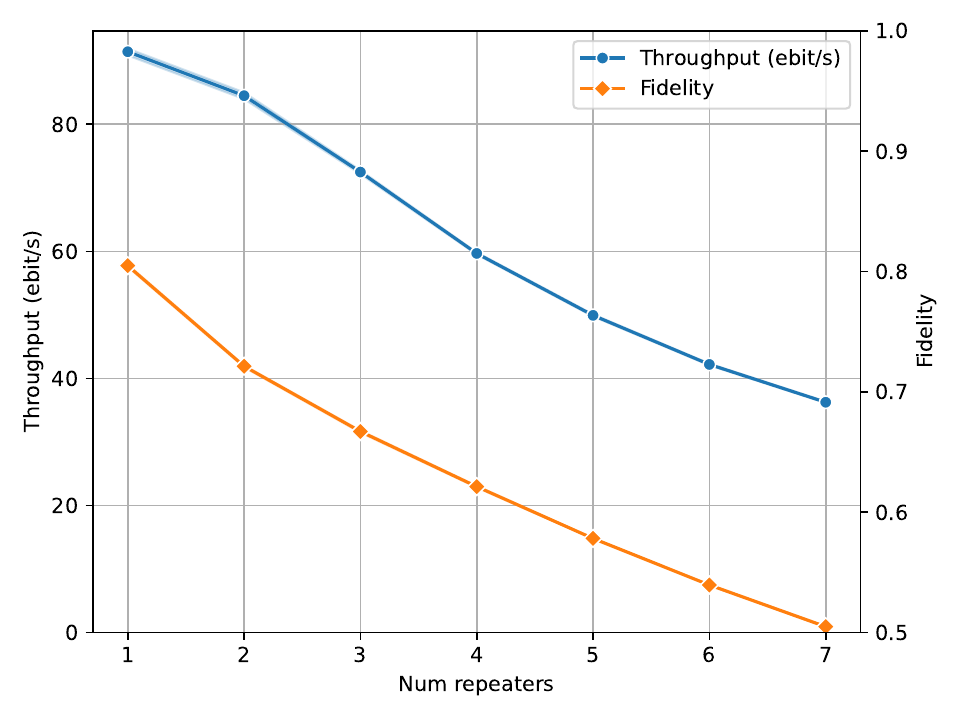}
\vspace{-1.5em}
\caption{Long-distance regime. Throughput and fidelity achieved by \hopper{} with increasing path length, a memory size of 50 qubits, and 30 applications.}
\label{fig:008-chain-perfect-varlen-scalar-throughput}
\vspace{-1.5em}
\end{figure}

In \Cref{fig:008-chain-perfect-varlen-scalar-throughput} we show the throughput and fidelity with \hopper, with an intermediate memory size of 50~qubits, 30 applications, and varying path size.
As expected, both the metrics decrease, but they do so gracefully, thanks to the use of quantum memories, the continuous asynchronous generation of local entanglement, and the \hopper{} protocol, which is designed to make the best of these features.
This is in stark contrast with alternative technologies for quantum networking, e.g., with entanglement swapping realized with linear optics components, where the rate decays exponentially with the path size, unless an $n$-partite entanglement is used, with $n > 2$~\cite{patil_entanglement_2022}.

\begin{figure}[tbp!]
\centering
\includegraphics[width=\columnwidth]{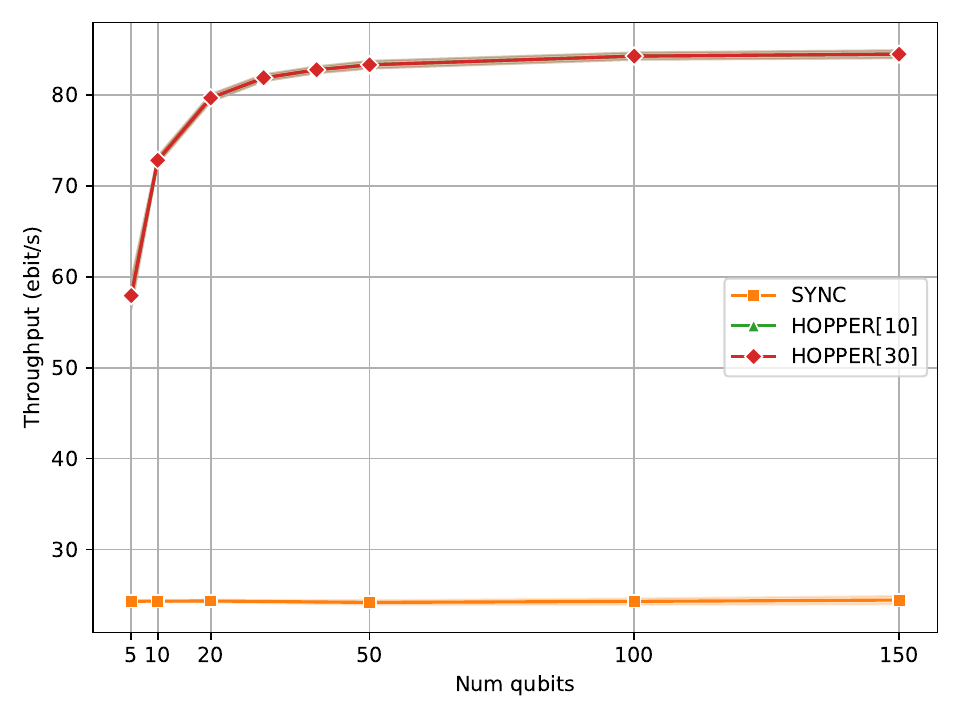}
\vspace{-1.5em}
\caption{Short-distance regime. Comparison of throughput between \sync{} and \hopper, the latter with 10 vs.\ 30 applications.}
\label{fig:007-chain-mini-compare-throughput}
\vspace{-1.5em}
\end{figure}

We conclude with the comparison of \sync{} and \hopper{} in a \textit{short-distance regime}, where the nodes in a 3-repeater path are just 5~m apart, for which we have used $\Gamma = 0.01$~Hz.
The throughput is reported in \Cref{fig:007-chain-mini-compare-throughput}.
As can be seen, the \sync{} curve is almost flat.
This is because, in this regime, the time slot duration is determined essentially by the local entanglement phase (in the long-distance regime, there was a trade-off with the signaling latency), and the link rate has to be divided proportionally between the memory cells: a memory of doubled size means a time slot of double duration, hence generating the same ebits/s.
On the other hand, \hopper{} can exploit the increasing memory size, and reaches, again, a plateau of about 84\%, like in the long-distance regime.
There is no noticeable difference in this scenario between 10 and 30 applications (or even smaller/higher numbers, not shown), because the classical latencies are so small that a single application can already fill the ``pipe'' by itself.

\section{Related work}\label{sec:soa}

We now briefly review some relevant studies.
As already mentioned, \ac{QNP} was proposed in the seminal work~\cite{kozlowski_designing_2020}, which mentioned the possibility of sharing links among applications using \textit{virtual circuits} in an ATM-like manner; in this paper, we have provided a practical solution to achieve the same objective, even if we kept in mind a TCP/IP-oriented design, with ebits identified by a five-tuple with node addresses and ports.
In \cite{soon_performance_2024}, the authors aimed at optimizing network performance along heterogeneous paths in a quantum network; the work has provided some key insights for the design of \hopper, including the grouping of memory qubits per link.
In \cite{beauchamp_modular_2025}, the authors elaborated on the possibility of adopting a synchronous approach using variable-length time slots and scheduling dynamically local entanglement with \ac{EDF}; however, this design is only feasible in small-scale networks, due to the complex signaling required, while with \hopper{} we aspired to create a protocol for generic quantum networks.
In \cite{yang_asynchronous_2024}, the authors study synchronous vs.\ asynchronous routing protocols from a different perspective than we do here: they address the problem of dynamically finding the destination node in a network where nodes are not aware of the topology beyond their neighbors.
On the other hand, \hopper{} defines the protocol to realize a hop-by-hop end-to-end entanglement as fast as possible, assuming that the path is known. Thus, it is a complementary contribution to \cite{yang_asynchronous_2024}, with which it would be interesting to study how to possibly integrate.
Likewise, the ebit establishment procedure is just a cog in the Quantum Network machine, which can be studied in cooperation with other important functions.
For instance, transport layer likely plays an important role in the overall performance, and it has clear relations with underlying networking protocols such as \hopper{}, especially for what concerns possible rate adaptation systems and congestion control/reliability, both of which have been already studied under the assumption of asynchronous operation at nodes, in \cite{11311584} and \cite{zhao_asynchronous_2024}, respectively.

\section{Conclusions}\label{sec:conclusions}

In this paper, we have proposed \hopper, a protocol for quantum networks designed to establish end-to-end entanglement between qubits at remote nodes by means of a hop-by-hop procedure not relying on node synchronization nor a central controller.
The protocol allows multiplexing of concurrent applications on the same path, and it is designed to minimize the classical signaling time, to reduce the negative impact of decoherence of qubits in the nodes' quantum memories.
We have shown via simulation that \hopper{} is superior, in terms of both throughput and end-to-end fidelity, to a synchronous protocol with pre-configured path allocation.
Furthermore, \hopper{} retains its property in short-distance and long-distance regimes alike.
The simulations have been performed along a single path and with a simple model for applications, quantum links, and memories, with research underway to expand to more complex and realistic use cases.
Furthermore, in this work, we have compared a centralized synchronous baseline against \hopper, which is decentralized and asynchronous, leaving unaddressed the mixed cases.

\begin{acronym}
  \acro{3GPP}{Third Generation Partnership Project}
  \acro{5G-PPP}{5G Public Private Partnership}
  \acro{5GC}{5G Core}
  \acro{AA}{Authentication and Authorization}
  \acro{ADF}{Azure Durable Function}
  \acro{AI}{Artificial Intelligence}
  \acro{API}{Application Programming Interface}
  \acro{AP}{Access Point}
  \acro{AR}{Augmented Reality}
  \acro{BGP}{Border Gateway Protocol}
  \acro{BSP}{Bulk Synchronous Parallel}
  \acro{BS}{Base Station}
  \acro{BSM}{Bell State Measurement}
  \acro{CDF}{Cumulative Distribution Function}
  \acro{CFS}{Customer Facing Service}
  \acro{CPU}{Central Processing Unit}
  \acro{DAG}{Directed Acyclic Graph}
  \acro{DHT}{Distributed Hash Table}
  \acro{DNS}{Domain Name System}
  \acro{EAS}{Edge Application Server}
  \acro{ECSP}{Edge Computing Service Provider}
  \acro{EDF}{Earliest Deadline First}
  \acro{EDN}{Edge Data Network}
  \acro{EPR}{Einstein-Podolsky-Rosen}
  \acro{EPSG}{Entangled Photon Source Generator}
  \acro{ETSI}{European Telecommunications Standards Institute}
  \acro{FCFS}{First Come First Serve}
  \acro{FSM}{Finite State Machine}
  \acro{FaaS}{Function as a Service}
  \acro{GHZ}{Greenberger-Horne-Zeilinger}
  \acro{GPU}{Graphics Processing Unit}
  \acro{HTML}{HyperText Markup Language}
  \acro{HTTP}{Hyper-Text Transfer Protocol}
  \acro{ICN}{Information-Centric Networking}
  \acro{ICT}{Information and Communication Technologies}
  \acro{IETF}{Internet Engineering Task Force}
  \acro{IIoT}{Industrial Internet of Things}
  \acro{ILP}{Integer Linear Programming}
  \acro{IPP}{Interrupted Poisson Process}
  \acro{IP}{Internet Protocol}
  \acro{ISG}{Industry Specification Group}
  \acro{ITS}{Intelligent Transportation System}
  \acro{ITU}{International Telecommunication Union}
  \acro{IT}{Information Technology}
  \acro{IaaS}{Infrastructure as a Service}
  \acro{IoT}{Internet of Things}
  \acro{JSON}{JavaScript Object Notation}
  \acro{K8s}{Kubernetes}
  \acro{KPI}{Key Performance Indicator}
  \acro{KVS}{Key-Value Store}
  \acro{LEO}{Low Earth Orbit}
  \acro{LCM}{Life Cycle Management}
  \acro{LOCC}{Local Operations and Classical Communication}
  \acro{LL}{Link Layer}
  \acro{LTE}{Long Term Evolution}
  \acro{MAC}{Medium Access Layer}
  \acro{MBWA}{Mobile Broadband Wireless Access}
  \acro{MCC}{Mobile Cloud Computing}
  \acro{MEC}{Multi-access Edge Computing}
  \acro{MEH}{Mobile Edge Host}
  \acro{MEPM}{Mobile Edge Platform Manager}
  \acro{MEP}{Mobile Edge Platform}
  \acro{ME}{Mobile Edge}
  \acro{ML}{Machine Learning}
  \acro{MNO}{Mobile Network Operator}
  \acro{NAT}{Network Address Translation}
  \acro{NFV}{Network Function Virtualization}
  \acro{NFaaS}{Named Function as a Service}
  \acro{NIC}{Network Interface Card}
  \acro{NN}{Neural Network}
  \acro{OSPF}{Open Shortest Path First}
  \acro{OSS}{Operations Support System}
  \acro{OS}{Operating System}
  \acro{OWC}{OpenWhisk Controller}
  \acro{PKI}{Public Key Infrastructure}
  \acro{PMF}{Probability Mass Function}
  \acro{PU}{Processing Unit}
  \acro{PaaS}{Platform as a Service}
  \acro{PoA}{Point of Attachment}
  \acro{QEC}{Quantum Error Correction}
  \acro{QKD}{Quantum Key Distribution}
  \acro{QNP}{Quantum Network Protocol}
  \acro{QoE}{Quality of Experience}
  \acro{QoS}{Quality of Service}
  \acro{QPU}{Quantum Processing Unit}
  \acro{RAN}{Radio Access Network}
  \acro{RPC}{Remote Procedure Call}
  \acro{RR}{Round Robin}
  \acro{RSU}{Road Side Unit}
  \acro{RTT}{Round Trip Time}
  \acro{SBC}{Single-Board Computer}
  \acro{SDN}{Software Defined Networking}
  \acro{SJF}{Shortest Job First}
  \acro{SLA}{Service Level Agreement}
  \acro{SMP}{Symmetric Multiprocessing}
  \acro{SoC}{System on Chip}
  \acro{SRPT}{Shortest Remaining Processing Time}
  \acro{SPT}{Shortest Processing Time}
  \acro{STL}{Standard Template Library}
  \acro{SaaS}{Software as a Service}
  \acro{TCP}{Transmission Control Protocol}
  \acro{TPU}{Tensor Processing Unit}
  \acro{TSN}{Time-Sensitive Networking}
  \acro{UDP}{User Datagram Protocol}
  \acro{UE}{User Equipment}
  \acro{UPF}{User Plane Function}
  \acro{URI}{Uniform Resource Identifier}
  \acro{URL}{Uniform Resource Locator}
  \acro{UT}{User Terminal}
  \acro{VANET}{Vehicular Ad-hoc Network}
  \acro{VIM}{Virtual Infrastructure Manager}
  \acro{VR}{Virtual Reality}
  \acro{VM}{Virtual Machine}
  \acro{VNF}{Virtual Network Function}
  \acro{WLAN}{Wireless Local Area Network}
  \acro{WMN}{Wireless Mesh Network}
  \acro{WRR}{Weighted Round Robin}
  \acro{YAML}{YAML Ain't Markup Language}
\end{acronym}





\end{document}